\documentclass[sigconf,screen]{acmart}
\AtBeginDocument{%
  }

\renewcommand\footnotetextcopyrightpermission[1]{} % removes footnote with conference 

\usepackage{multirow}
\usepackage{subcaption}
\usepackage[most]{tcolorbox}
\newcommand{\find}[1]{
\begin{tcolorbox}[tile,size=fbox,boxsep=2mm,boxrule=0pt,top=0pt,bottom=0pt,
borderline west={1mm}{0pt}{blue!50!white},colback=blue!5!white]
\em #1
\end{tcolorbox}
}

\begin{document}

\title{MultiKernelBench: A Multi-Platform Benchmark for Kernel Generation}

\author{Zhongzhen Wen}
\orcid{0009-0005-6737-9953}
\affiliation{%
  \institution{State Key Lab for Novel Software Technology, Nanjing University}
  \city{Nanjing}
  \country{China}
}
\email{wenzhongzhen@smail.nju.edu.cn}

\author{Yinghui Zhang}
\affiliation{%
  \institution{State Key Lab for Novel Software Technology, Nanjing University}
  \city{Nanjing}
  \country{China}
}
\email{zhangyh.halo@smail.nju.edu.cn}

\author{Zhong Li}
\orcid{0000-0002-3849-3416}
\affiliation{%
  \institution{State Key Lab for Novel Software Technology, Nanjing University}
  \city{Nanjing}
  \country{China}
}
\email{lizhong@nju.edu.cn}

\author{Zhongxin Liu}
\orcid{0000-0002-1981-1626}
\affiliation{%
  \institution{The State Key Laboratory of Blockchain and Data Security, Zhejiang University}
  \city{Hangzhou}
  \country{China}
}
\email{liu_zx@zju.edu.cn}

\author{Linna Xie}
\orcid{0000-0003-4163-8994}
\affiliation{%
  \institution{State Key Lab for Novel Software Technology, Nanjing University}
  \city{Nanjing}
  \country{China}
}
\email{xieln@smail.nju.edu.cn}

\author{Tian Zhang}
\orcid{0000-0003-0104-2731}
\affiliation{%
  \institution{State Key Lab for Novel Software Technology, Nanjing University}
  \city{Nanjing}
  \country{China}}
\email{ztluck@nju.edu.cn}


\begin{abstract}
The automatic generation of deep learning (DL) kernels using large language models (LLMs) has emerged as a promising approach to reduce the manual effort and hardware-specific expertise required for writing high-performance operator implementations. However, existing benchmarks for evaluating LLMs in this domain suffer from limited hardware support, coarse-grained kernel categorization, and imbalanced task coverage. To address these limitations, we introduce MultiKernelBench, the first comprehensive, multi-platform benchmark for LLM-based DL kernel generation. MultiKernelBench spans 285 tasks across 14 well-defined kernel categories and supports three major hardware platforms: Nvidia GPUs, Huawei NPUs, and Google TPUs. To enable future extensibility, we design a modular backend abstraction layer that decouples platform-specific logic from the core benchmarking infrastructure, allowing easy integration of new hardware platforms. We further propose a simple yet effective category-aware one-shot prompting method that improves generation quality by providing in-category exemplars. Through systematic evaluations of seven state-of-the-art LLMs, we reveal significant variation in task difficulty, poor generalization to platforms with less training exposure, and the effectiveness of targeted prompting strategies. MultiKernelBench is publicly available at \url{https://github.com/wzzll123/MultiKernelBench}.
\end{abstract}





\maketitle

\section{Introduction}
The execution of deep learning (DL) models fundamentally depends on the execution of DL kernels, which are implemented using low-level parallel programming languages such as CUDA for Nvidia GPUs and AscendC for Huawei NPUs. Writing correct and high-performance kernels is a time-consuming process that demands deep hardware-specific expertise. 
Recent advances in Large Language Models (LLMs) offer a promising approach for automatically generating such kernels.
Particularly, several kernel benchmarks~\cite{ouyang2025kernelbench,li2025tritonbenchbenchmarkinglargelanguage} have been proposed to evaluate the capabilities of LLMs and support future research in this direction.

\textbf{Research gaps} - However, our analysis of existing benchmarks reveals three major limitations of them: \textbf{(1) Limited platform support.} Most existing benchmarks focus solely on Nvidia GPUs, neglecting other widely-used platforms such as Huawei NPUs and Google TPUs. These platforms also require kernel development, and the unique challenges they present remain largely unexplored. Moreover, due to the tightly coupled design of existing benchmarks~\cite{ouyang2025kernelbench,li2025tritonbenchbenchmarkinglargelanguage}—binding build and evaluation procedures specifically to the Nvidia GPU ecosystem—extending them to support multiple platforms is non-trivial. \textbf{(2) Lack of detailed categorization.} Existing benchmarks either do not categorize kernels at all~\cite{li2025tritonbenchbenchmarkinglargelanguage} or use overly coarse categorizations (e.g., three or four levels of complexity)~\cite{ouyang2025kernelbench}. This lack of granularity is problematic, as evidenced by the wide variation in LLM-generated kernel performance even within the same level. For example, in our experiment, DeepSeek-V3 achieved a 0\% correctness rate on CUDA kernels for the convolution category, while achieving over 80\% correctness for the activation category—both classified under the same difficulty level in the existing benchmark~\cite{ouyang2025kernelbench}. \textbf{(3) Imbalanced kernel distribution.} While most categories contain a reasonable number of kernels, some categories have very few, and certain important categories, such as optimizer kernels, are missing altogether, resulting in an uneven and insufficient evaluation landscape.
% some categories contain very few kernels, while others are missing entirely, leading to an uneven and insufficient evaluation landscape.

\textbf{MultiKernelBench} – To address these limitations, we introduce MultiKernelBench, a comprehensive, multi-platform benchmark suite for evaluating LLMs in the context of deep learning kernel generation. Unlike prior benchmarks that focus solely on Nvidia GPUs, MultiKernelBench supports diverse hardware platforms—including Nvidia GPUs (CUDA), Huawei NPUs (AscendC), and Google TPUs (Pallas)—enabling a more realistic and broadly applicable evaluation. We categorize DL kernels by their functional characteristics and extend the task set from KernelBench~\cite{ouyang2025kernelbench}, adding missing categories and augmenting existing ones. In total, MultiKernelBench includes 285 kernel generation tasks across 14 well-defined categories, supporting fine-grained, platform-agnostic evaluation. To ensure extensibility, we implement a modular backend abstraction layer that cleanly separates platform-specific logic from the core evaluation pipeline. New hardware targets can be added by implementing a unified backend interface for device setup, compilation, execution, and resource management—without altering core logic.
\begin{figure*}[t]
  \centering

  % ---------- LEFT COLUMN (stacked) ----------
  \begin{subfigure}[b]{0.47\textwidth}
    \centering

    % first picture + its own little caption
    \subcaptionbox{CUDA Kernel\label{fig:cuda_add}}{%
      \includegraphics[width=\linewidth]{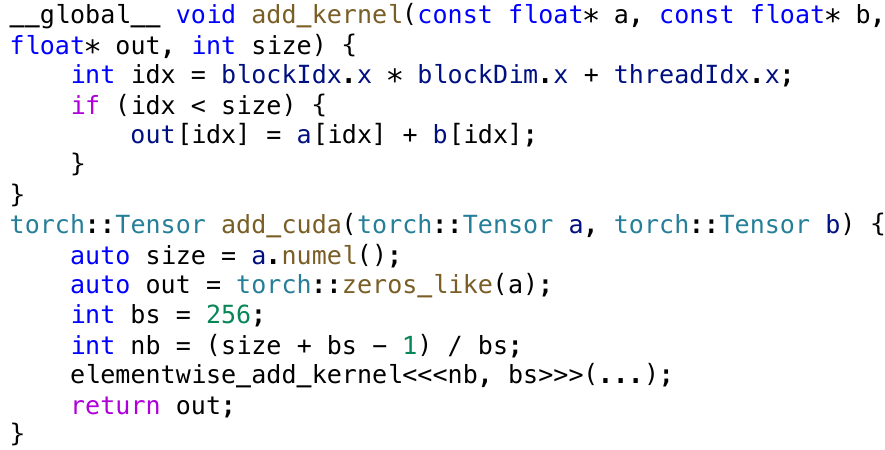}}%
    \vspace{0.5em} % Adjust this value as needed
    % second picture + its own little caption
    \subcaptionbox{Pallas Kernel\label{fig:pallas_add}}{%
      \includegraphics[width=\linewidth]{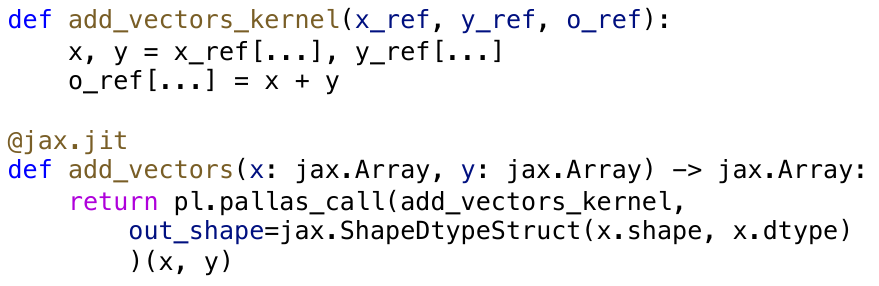}}
  \end{subfigure}%
  \hfill
  % ---------- RIGHT COLUMN ----------
  \begin{subfigure}[b]{0.49\textwidth}   % same width!
    \centering
    \subcaptionbox{AscendC Kernel\label{fig:ascendc_add}}{%
      \includegraphics[width=\linewidth]{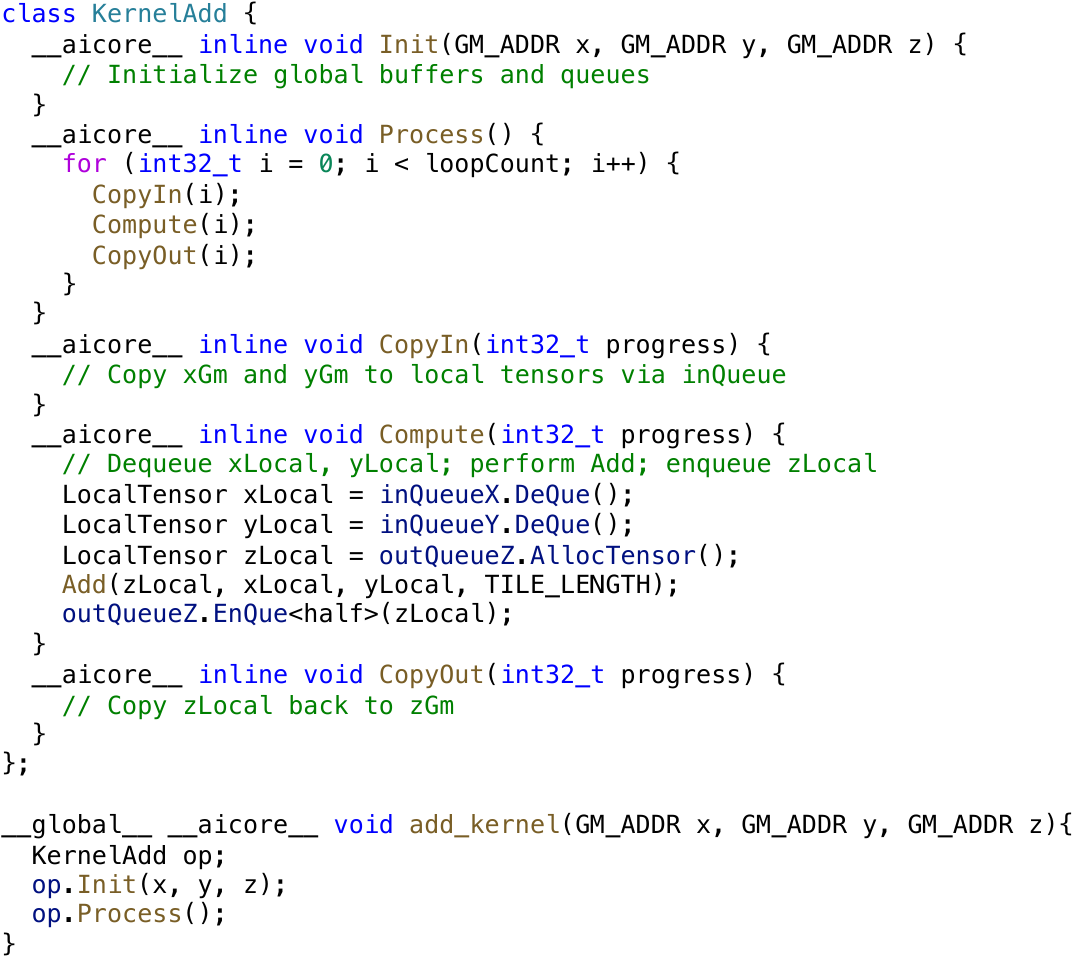}}
  \end{subfigure}

  \caption{Vector-addition kernels written for three platforms.}
  \label{fig:add_kernels}
\end{figure*}

\textbf{Evaluations} – We further evaluate a set of LLMs under various settings using MultiKernelBench. We select seven diverse LLMs—including DeepSeek, Qwen, GPT-4o, and Claude—covering a range of scales from 32B to 681B parameters, and spanning both base and reasoning models. To systematically assess kernel generation, we report three key metrics: \textit{Compilation@k}, \textit{Pass@k}, and \textit{$SpeedUp_\alpha@k$}, which capture performance from basic compilation success to runtime speedup. We design a unified prompt template compatible across multiple platforms and explore different prompting strategies. In the default setting, we use the add task as a one-shot example to demonstrate basic syntax and formatting. Additionally, we experiment with selecting one-shot examples from the same category as the target tasks, helping LLMs leverage relevant strategies and domain-specific knowledge. Following evaluation, we conduct extensive analysis, including failure diagnosis, category-wise assessments, and speedup case studies. Our experiments reveal several key insights, for example, category-aware one-shot prompting notably improves performance on platforms with limited training exposure, and LLMs occasionally discover meaningful optimizations, such as kernel fusion and exploiting data sparsity.
% (1) LLM performance varies significantly across platforms, highlighting gaps in training corpus coverage—models perform best on CUDA, but often fail on AscendC and Pallas.
% (2) Task performance differs greatly by category, revealing large variation in difficulty across kernel types.
% (3) 
% (4) LLMs occasionally discover meaningful optimizations, including kernel fusion and exploiting data sparsity.

In summary, the paper makes the following contributions:
\begin{itemize}
\item We introduce MultiKernelBench, the first multi-platform benchmark for kernel generation, spanning GPUs, NPUs, and TPUs with 285 tasks in 14 categories.

\item We introduce an extensible backend abstraction layer that decouples platform-specific logic from the core evaluation pipeline. New hardware platforms can be added by implementing a unified backend interface, enabling easy integration and maintenance across diverse platforms.

\item We propose a simple yet effective \emph{category‑aware one‑shot} strategy that selects an exemplar kernel from the same category as the target task. This improves LLM performance by providing domain-specific patterns, especially on platforms with limited representation in LLM training data.
\item We conduct extensive experiments on MultiKernelBench using seven LLMs under various settings. Our analysis reveals key insights on platform sensitivity, category-wise difficulty, and the benefits of category-aware prompting.
\end{itemize}

\section{Background}

\subsection{Hardware for Deep Learning Computing}
Larger training datasets and an increased number of model parameters make deep learning models more powerful—but at the cost of significantly higher computational and storage demands. Traditional CPU architectures are no longer sufficient to meet these intensive requirements. To address this challenge, several custom hardware solutions have been developed to accelerate deep learning training and inference. Among them, Nvidia’s GPU, Google’s TPU and Huawei’s NPU  stand out as representative hardware, each with distinct architectures and design philosophies.

Nvidia’s GPU (Graphics Processing Unit) was originally developed for graphics rendering. Modern GPUs feature thousands of Arithmetic Logic Units within a single processor, allowing DL workloads to run in parallel with high efficiency. However, they still rely heavily on memory and register access to store intermediate variables during operations like matrix multiplication. 
To further accelerate such workloads, Nvidia introduced Tensor Cores~\cite{tensor-cores}—specialized hardware units optimized for fast, mixed-precision matrix operations that are critical in DL.

Google’s TPU (Tensor Processing Unit), on the other hand, is purpose-built for deep neural network computations. TPUs consist of thousands of multiply-accumulate units that are directly interconnected to form a large, physical matrix. This architecture eliminates the need for memory access during matrix multiplication, greatly enhancing performance for specific workloads. However, TPUs are less suited for tasks that involve frequent branching or element-wise operations~\cite{tpu-arch}.

Huawei’s NPU (Neural Processing Unit) adopts a heterogeneous architecture that integrates scalar, vector, and cube computing units~\cite{Ascend, Squeezing_Ascend}. Scalar units manage logical control, vector units handle element-wise and vector-related operations, and cube units specialize in matrix computations. This design enables the three types of units to operate in parallel, significantly improving computational efficiency through specialization and coordination.

\subsection{Deep Learning Kernels}

To harness the power of deep learning accelerators, developers need to write programs using specific languages or libraries tailored for the underlying hardware. These low-level programs—often referred to as kernels—directly implement computational primitives such as matrix multiplication, convolution, and activation functions. Writing efficient kernels requires careful attention to memory hierarchy, parallelism, and hardware-specific features.

CUDA is Nvidia’s parallel computing platform and programming model, which allows developers to write GPU-accelerated code. CUDA exposes the massive parallelism of GPUs by allowing fine-grained control over threads, memory (global, shared, and local), and execution hierarchy (grids and blocks). An example CUDA kernel for vector addition is shown in Figure~\ref{fig:cuda_add}. It performs element-wise addition on two input arrays in parallel using thread indices.

AscendC is Huawei's low-level kernel programming model designed for the Ascend NPU. It offers direct access to the hardware’s heterogeneous compute units—scalar, vector, and cube—and allows precise control over memory transfers and execution queues. An example AscendC kernel for vector addition is shown in Figure~\ref{fig:ascendc_add}. It uses the AscendC API function \verb|Add| to perform vector addition on a tile using the vector units. The kernel coordinates computation, data movement, memory allocation, and task synchronization through explicit API calls.

Pallas is a high-level kernel programming interface built on top of the JAX ecosystem. It allows users to define custom TPU kernels using a Python syntax, abstracting many complexities of parallel programming. JAX transformations are then applied to lower these simple, high-level descriptions into efficient low-level code. An example Pallas kernel for vector addition is shown in Figure~\ref{fig:pallas_add}. The kernel is invoked within a JAX computation using the \verb|pallas_call| higher-order function. Among the three examples, the Pallas kernel is the most concise, reflecting its simplicity and ease of use.
\section{MultiKernelBench}

\subsection{Benchmark Overview}
\begin{figure*}[t]
  \centering
  \includegraphics[width=0.85\textwidth]{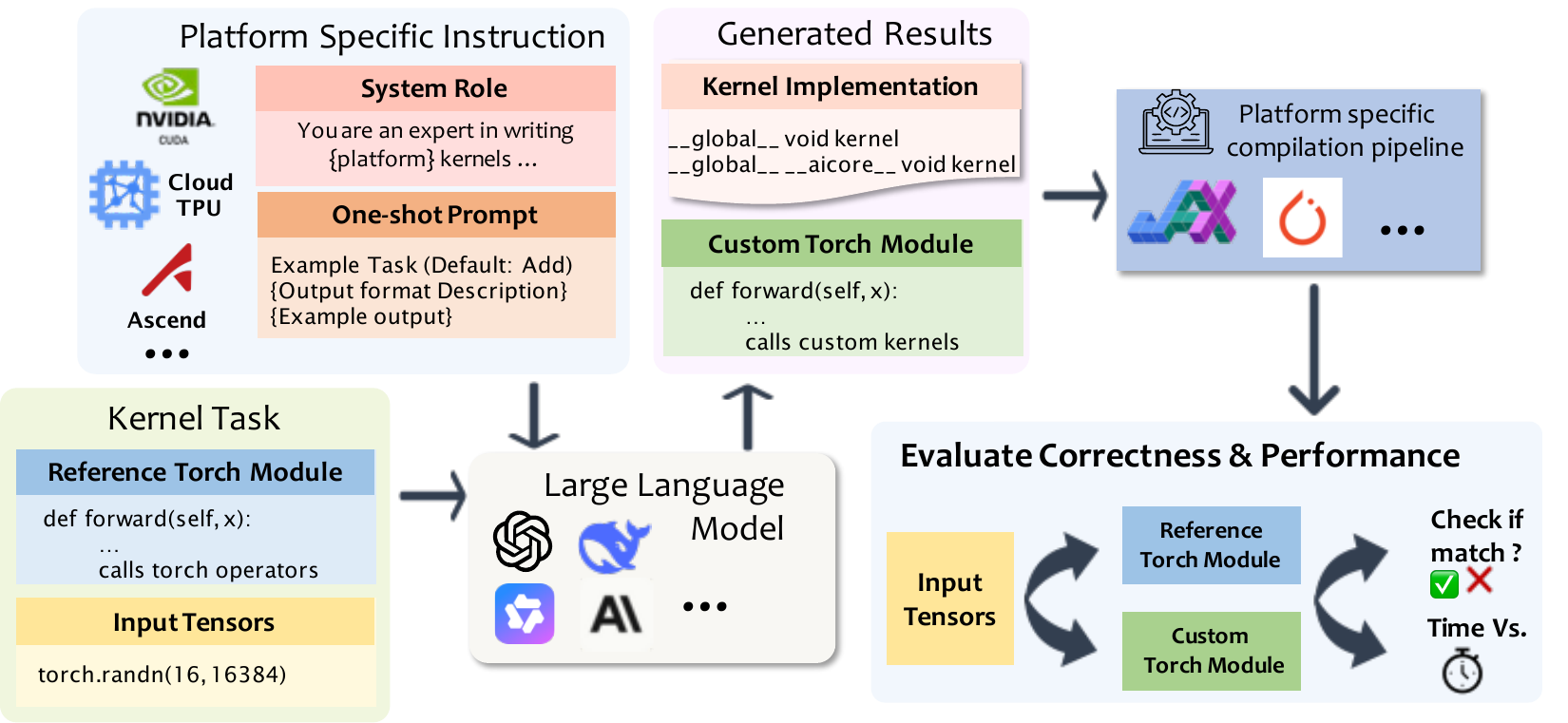}
  \caption{Framework of MultikernelBench.}
  \label{fig:overview}
\end{figure*}
Figure~\ref{fig:overview} illustrates the overall framework of MultiKernelBench. The benchmark defines kernel tasks using reference PyTorch modules, where the forward method invokes official operators, alongside input tensors that specify the data the kernel must process. Each kernel task, combined with platform-specific instructions that define the system role and provide a one-shot prompt specifying the desired output format, is given to the LLM. The LLM generates two components: a custom kernel implementation for the target platform and custom PyTorch module code that invokes the custom kernel. After generation, the benchmark includes a build pipeline that compiles all LLM-generated outputs into an executable PyTorch module, aiming to match the functionality of the original module while improving performance. Finally, both the generated and reference modules are executed with the same input tensors to verify correctness and evaluate performance.

\subsection{Input Prompt and Target Output}\label{sec:prompt}
\noindent\textbf{Input. }We design a unified prompt template for kernel generation across multiple platforms. As illustrated in Figure~\ref{fig:overview}, each prompt consists of two parts: (1) a platform-specific instruction that defines the system role, describes the task, and provides a one-shot example illustrating the required output; and (2) the target kernel task, including the PyTorch module to transform and its input tensors. The second part remains consistent across platforms, while platform-specific differences are primarily reflected in the one-shot example. Each example begins with a kernel task—typically add—followed by a description of the platform-specific format and the output, helping the LLM learn both domain conventions and the required output structure.

\noindent\textbf{Output. }For CUDA and Pallas, the expected output is self-contained Python code. In the CUDA case, this includes C++ kernel code embedded as Python strings, with a PyTorch module invoking the kernel via \texttt{cpp\_extension}~\cite{cpp_extension} during its forward function. For Pallas, the output consists of a Python-defined kernel function wrapped with \texttt{pallas\_call}, along with JAX and PyTorch integration code that compiles the kernel and exposes it as a PyTorch-compatible module. For AscendC, the output includes structured components such as host code, tiling configuration, device code, and the PyTorch module. This output format is designed to support future compilation, as discussed in Section~\ref{sec:building}.
\subsection{Kernel Tasks}
The existing benchmark, KernelBench~\cite{ouyang2025kernelbench}, provides a set of 250 kernel tasks designed for evaluating large language models. These tasks are grouped into three difficulty levels and cover a range of topics, including convolutions and matrix multiplication with various input shapes. While KernelBench offers a solid foundation for kernel task evaluation, it exhibits several limitations. First, its classification scheme—based solely on three coarse difficulty levels—lacks the granularity required for more detailed analysis. Second, the benchmark omits several important categories of kernel operations, limiting its comprehensiveness.

To address these shortcomings, we propose a new taxonomy grounded in the functional semantics of kernel operations rather than relying solely on difficulty levels. We followed a systematic procedure to construct a broader and more representative classification of kernel tasks.
(1) Initial Categorization of Existing Tasks: We first categorized the 250 kernel tasks from KernelBench based on computational patterns (such as reductions and matrix multiplications) and functional roles within typical DL pipelines (such as activation functions and normalization layers).
(2) Gap Identification via Kernel Library Review: We then reviewed Pytorch native CUDA kernel library~\cite{aten-codebase}, assessing whether each kernel could be mapped to an existing category. Kernels that did not fit were collected as candidates for new categories.
(3) Integration and Refinement: Candidate kernels were systematically categorized, and category definitions were refined through discussions among three kernel developers, each with three years of experience, to ensure completeness and consistency.
(4) Pruning Sparse and Low-Impact Categories: Categories with few representative tasks or limited relevance to common DL scenarios were removed, based on consensus among the three developers. For instance, comparison-related kernels were excluded due to their simplicity and negligible impact on LLM performance evaluation.

This process resulted in a set of 14 functional categories capturing a broad and representative range of kernel operations, including Resize, Reduction, Normalization, and Indexing. Compared to KernelBench, we introduce three new categories—Broadcast, Resize, and Optimizer—to cover previously unrepresented operations, and we expand the Indexing category, which originally included only two tasks, to better reflect real-world usage. New tasks are selected from existing kernel libraries~\cite{aten-codebase}, then refined and manually verified to ensure practical relevance and comprehensive coverage. Table~\ref{table:category} summarizes these categories, along with representative tasks and the number of tasks per category, providing a more fine-grained and functional framework for evaluating LLM performance in DL kernel generation.

% To address these shortcomings, we propose a new categorization scheme grounded in the functional semantics of kernel operations, rather than relying solely on coarse difficulty levels. In this scheme, tasks are grouped based on their underlying computational patterns and the functional roles they serve in typical machine learning pipelines. Representative categories include Resize, Reduction, Normalization, and Indexing. This fine-grained organization enables more targeted and interpretable evaluations of LLM performance, helping to reveal specific strengths and weaknesses in the model's understanding of various kernel behaviors.

% Table~\ref{table:category} presents the 14 categories used in our benchmark, along with representative tasks and the number of tasks per category. Compared to KernelBench, we introduce three new categories - Broadcast, Resize, and Optimizer - to cover operations that were previously unrepresented. Additionally, we expand the Index category, which originally included only two tasks, to better reflect the diversity of indexing operations encountered in real-world applications.
\begin{table}[t]\small
\caption{Categorization and counts of kernels.}\label{table:category}
\centering
\begin{tabular}{lll}
\hline
\textbf{Categories}  & \textbf{Representative} & \textbf{\#Tasks}    \\
\hline
Activation         & \texttt{relu}, \texttt{gelu}                        & 15 \\
Broadcast          & \texttt{bias\_add}           & 10 \\
Convolution        & \texttt{conv2d}    & 34 \\
Full Architecture  & \texttt{resnet18} & 50 \\
Fusion             & \texttt{fused\_matmul\_bias} & 100 \\
Loss               & \texttt{cross\_entropy}, \texttt{mse}               & 7 \\
Math               & \texttt{multiply}                          & 6 \\
Matrix Multiply    & \texttt{sgemm}, \texttt{bmm}                         & 17 \\
Normalization      & \texttt{batchnorm}, \texttt{layernorm}              & 8 \\
Optimizer          & \texttt{adam\_update}, \texttt{sgd\_momentum}       & 5 \\
Pooling            & \texttt{maxpool2d}, \texttt{avgpool2d}              & 6 \\
Index              & \texttt{gather}, \texttt{scatter\_update}           & 12 \\
Resize             & \texttt{bilinear\_resize} & 10 \\
Reduce             & \texttt{reduce\_sum}, \texttt{reduce\_max}          & 5 \\

\hline
Total&& 285\\
\hline
\end{tabular}
\end{table}

\subsection{Platform-Specific Compilation}\label{sec:building}
After generation, MultiKernelBench transforms the LLM-generated text into an executable PyTorch module for further evaluation.

As discussed in Section~\ref{sec:prompt}, the generated output for Pallas and CUDA is already self-contained Python code. The compilation pipeline simply extracts the code from the text and dynamically executes it using \texttt{exec}~\cite{python-exec}. The actual build process, such as compiling CUDA source code with \texttt{nvcc}, is handled automatically by utilities like \texttt{cpp\_extension}~\cite{cpp_extension}. Additionally, for CUDA, device-side assertions should be enabled for failure analysis.

For AscendC, the build process is more involved. Unlike CUDA and Pallas, AscendC does not provide automatic utilities to compile and bind these components directly from Python. Therefore, we define the AscendC output format to include multiple components, as discussed in Section~\ref{sec:prompt}. MultiKernelBench uses glue scripts to organize these components into the required file structure, execute the AscendC compilation pipeline, and produce an executable PyTorch module.

\subsection{Evaluation Approach}
MultiKernelBench automatically evaluates the quality of generated kernels using three key metrics: compilation success, correctness, and performance.

\noindent \textbf{Compilation Success.} MultiKernelBench determines compilation success based on whether exceptions occur during the platform-specific build process, as described in Section~\ref{sec:building}. Since CUDA and AscendC use C/C++-like languages, more errors tend to be caught at compile time compared to Pallas, which is Python-based. This metric reflects the minimum requirement for understanding a platform’s syntax and language-specific constraints.

\noindent \textbf{Correctness. }To evaluate correctness, MultiKernelBench adopts a similar approach to prior work~\cite{ouyang2025kernelbench, mirage}, using randomized testing to verify kernel outputs. It generates $N$ random inputs and feeds them into both the LLM-generated module (producing $output_{llm}$) and a reference module (producing $output_{ref}$). The kernel is considered correct if it satisfies the condition:  $|output_{llm} - output_{ref}| < atol + rtol* |output_{ref}|$. Here, \textit{atol} (absolute tolerance) accounts for fixed small differences, and \textit{rtol} (relative tolerance) scales with the magnitude of the reference output to handle proportional errors. 

We follow prior work in setting the parameters: $N=5$, $atol=1e{-2}$, and $rtol=1e{-2}$. Previous studies~\cite{ouyang2025kernelbench} have observed that, on CUDA, five random tests often yield results consistent with running 100 tests. We reproduced this phenomenon on AscendC and Pallas, confirming that a small number of tests is generally sufficient for correctness evaluation. The relatively large tolerance values also accommodate potential precision-reducing optimizations, such as using BF16 or FP8.

\noindent \textbf{Performance. }MultiKernelBench measures performance using platform-specific methods. For CUDA and AscendC, we use synchronization markers~\cite{cuda-event,npu-event} placed before and after kernel execution to compute elapsed time between the two events. For Pallas, however, no such synchronization markers are available in the current interface. Instead, we use debugging tools to measure execution time. To ensure consistency, we conducted experiments on the same machine at different times and confirmed that the timing methods across all three platforms are stable, with controlled variance within a small range.

\subsection{Extensible Backend Abstraction for Multi-Platform Evaluation}
To enable kernel benchmarking across diverse hardware platforms, we introduce a modular backend architecture that cleanly decouples platform-specific logic from the core evaluation pipeline.

Each platform is supported by a dedicated subclass of a unified \texttt{Backend} interface. This interface encapsulates the essential responsibilities of: (1) device initialization and hardware identification, (2) kernel code compilation and execution, (3) correctness verification and performance measurement, and (4) resource cleanup and memory management.

The benchmark framework dynamically loads the appropriate backend at runtime based on the specified platform name. New hardware platforms can be supported by simply implementing and registering a new backend class using the \texttt{@register\_backend(<platform>)} decorator—without modifying any core evaluation logic. This plugin-based design ensures that the system remains extensible, maintainable, and easily adaptable to emerging accelerator architectures with minimal engineering effort.

\section{EXPERIMENT}
In this section, we experiment with a set of LLMs configured in various ways and analyze the results. Specifically, we address the following three research questions:

\noindent \textbf{RQ1:} How do the evaluated LLMs perform when prompted with the add task as a one-shot example?

\noindent \textbf{RQ2:} How does performance vary across different task categories?

\noindent \textbf{RQ3:} Can in-category examples improve LLM-generated kernel correctness for platforms with limited training exposure?
\subsection{Methodology}
\begin{table}[h]\small
\caption{Specifications of hardware used in experiments}
\centering
\begin{tabular}{lll}
\hline
\textbf{Hardware} & \textbf{FP16 TFLOPS} & \textbf{HBM}  \\
\hline
NVIDIA L20 GPU & 59.35 & 48 GB  \\
Huawei Ascend 910B2 & 400 & 64 GB  \\
Google TPU v2-8& 180 & 64 GB \\
\hline
\end{tabular}
\label{tab:hardware_specs}
\end{table}
\noindent\textbf{Hardware. } Table~\ref{tab:hardware_specs} summarizes the specifications of the three hardware platforms used in our experiments: the NVIDIA L20 GPU, Huawei Ascend 910B2, and Google TPU v2-8. These platforms feature diverse architectures and capabilities tailored for large-scale AI workloads. The NVIDIA L20 GPU is based on the Ada Lovelace architecture, offering high performance for both training and inference tasks. The Huawei Ascend 910B2 is a high-performance AI processor built on the Da Vinci architecture, optimized for efficient matrix computation. The Google TPU v2-8 is a custom ASIC specifically designed for deep learning, delivering high throughput for tensor operations and seamlessly integrated into the Google Cloud infrastructure.

\begin{table}[h]\small
\caption{Studied Large Language Models}
\centering
\begin{tabular}{llll}
\hline
\textbf{Model} & \textbf{Type} & \textbf{Size} & \textbf{Time}  \\
\hline

DeepSeek-V3-0324 & Non-reasoning & 685B & 2025-03-24 \\
Qwen3-235B & Non-reasoning & 235B & 2025-04-29 \\
Qwen2.5-Coder & Non-reasoning & 32B & 2024-11-12 \\
GPT-4o & Non-reasoning & - & 2024-11-20 \\
Claude-Sonnet-4 & Non-reasoning & - & 2025-05-22 \\
DeepSeek-R1-0528 & Reasoning & 685B & 2025-05-28 \\
Qwen3-235B (think) & Reasoning & 235B & 2025-04-29 \\

\hline
\end{tabular}
\label{tab:llm}
\end{table}

\noindent\textbf{Studied LLMs. } Table~\ref{tab:llm} lists the seven large language models used in our experiments. These models differ in scale, design, and functional capabilities, covering a wide spectrum of use cases—from general-purpose language generation to advanced reasoning. DeepSeek-V3-0324 and DeepSeek-R1-0528 are the latest and most powerful open source models in the DeepSeek series, representing state-of-the-art performance in base and reasoning categories, respectively, each with 685B parameters. GPT-4o is one of the most powerful proprietary base models available. Claude-Sonnet-4 is a powerful model for programming, ranking first in programming token usage across models~\cite{openrouter-ranking}. Qwen2.5-Coder is a relatively compact 32B model trained primarily on code, serving as a representative of smaller, domain-specific LLMs. Qwen3-235B~\cite{qwen3} uniquely integrates two distinct operating modes—reasoning (“think”) and non-reasoning—within a single 235B-parameter model, enabling a direct comparison of reasoning augmentation on performance in various tasks.
\begin{table*}[htbp]\small
  \centering
  \caption{
    \textbf{LLM performance on 285 DL kernel tasks across CUDA, AscendC, and Pallas using \textit{add} as the one-shot example.} 
    We report compilation success (Comp@k), functional correctness (Pass@k), and speed-up ($SU_1$@k) under greedy (N=1) and sampling (N=5) decoding. "Total Pass" counts correct kernels across all platforms.
  }
  \label{tab:llm_kernel_stats}

  % ------------------- @1 Metrics -------------------
  \subcaption*{\textbf{Greedy Search N=1}}
\begin{tabular}{lccc ccc ccc c}
  \toprule
  \multirow{2}{*}{Models} 
    & \multicolumn{3}{c}{CUDA (\%)} 
    & \multicolumn{3}{c}{AscendC (\%)} 
    & \multicolumn{3}{c}{Pallas (\%)} 
    & \multirow{2}{*}{Total Pass (\#)} \\
  \cmidrule(lr){2-4} \cmidrule(lr){5-7} \cmidrule(lr){8-10}
    & Comp@1 & Pass@1 & $SU_1$@1
    & Comp@1 & Pass@1 & $SU_1$@1
    & Comp@1 & Pass@1 & $SU_1$@1
    & \\
  \midrule
  DeepSeek-V3-0324         & 74.7 & 21.4 & 6.3 & \textbf{10.2} & \textbf{2.5} & \textbf{1.1} & 70.9 & 4.6 & 4.2 & 81 \\
  Qwen3-235B               & 66.3 & 22.5 & 5.3 & 0.7 & 0.7 & 0.4 & 89.8 & 3.9 & 1.8 & 77 \\
  Qwen2.5-Coder-32B        & 66.0 & 15.8 & 3.5 & 2.8 & 1.8 & 0.7 & 94.7 & 4.6 & 3.2 & 63 \\
  GPT-4o                   & \textbf{97.5} & 23.2 & 5.3 & 2.5 & 1.8 & 0.4 & 97.2 & 6.3 & 2.8 & 89 \\
  Claude-Sonnet-4  & 92.3 & 47.0 & 20.4 & 5.3 & 2.1 & 0.4 & \textbf{99.6} & \textbf{8.4} & \textbf{7.7} & \textbf{164} \\
   % 17(pallas Pass@1)
  DeepSeek-R1-0528         & 75.4 & \textbf{52.6} & \textbf{26.0} & 1.4 & 1.4 & 0.0 & 81.4 & 2.8 & 1.4 & 162 \\
  Qwen3-235B (think)    & 79.3 & 44.2 & 19.3 & 0.7 & 0.7 & 0.0 & 93.0 & 7.4 & 7.0 & 149 \\
  \bottomrule
\end{tabular}

  \vspace{2em}

  % ------------------- @5 Metrics -------------------
  \subcaption*{\textbf{Sampling Search N=5}}
  \begin{tabular}{lccc ccc ccc c}
    \toprule
    \multirow{2}{*}{Models} 
      & \multicolumn{3}{c}{CUDA (\%)} 
      & \multicolumn{3}{c}{AscendC (\%)} 
      & \multicolumn{3}{c}{Pallas (\%)}
      & \multirow{2}{*}{Total Pass (\#)} \\
    \cmidrule(lr){2-4} \cmidrule(lr){5-7} \cmidrule(lr){8-10}
      & Comp@5 & Pass@5 & $SU_1$@5
      & Comp@5 & Pass@5 & $SU_1$@5
      & Comp@5 & Pass@5 & $SU_1$@5 \\
    \midrule
    DeepSeek-V3-0324         & 98.2 & 35.1 & 13.7 & 7.7 & 2.8 & 1.4
                            & 97.9  & 6.0  & 5.6 & 125 \\
    Qwen3-235B               & 97.2 & 43.9 & 13.0 & 3.9 & 2.1 & 1.4 
                            & 100.0  & 6.3  & 4.2 & 149 \\
    Qwen2.5-Coder-32B        & 94.7 & 31.9 & 11.9 & 4.2 & 2.5 & 1.1 
                            & 100.0 & 5.3  & 3.9 & 113 \\
    GPT-4o                   & \textbf{99.6} & 32.3 & 12.6 & 3.2 & 2.1 & 0.7 
                            & 100.0  & 9.8  & 8.8 & 126 \\
    Claude-Sonnet-4    & 97.9 & \textbf{55.8} & \textbf{26.0} & \textbf{8.1} & \textbf{3.9} & \textbf{2.1}  & \textbf{100.0}  & \textbf{10.5}  & \textbf{8.8} & \textbf{200} \\
    % The data corresponding to "total" is the statistical result of mean of Pallas(N=5) and does not yet include CUDA and AscendC for 2025/7/11.
    \bottomrule
  \end{tabular}

\end{table*}

\noindent\textbf{Reported Metrics.} In line with prior work on code generation~\cite{code_llama, DOMAINEVAL, origin_pass}, we assess functional correctness using the widely adopted \textit{Pass@k} metric. Under this metric, $k$ code samples are generated per problem, and a problem is considered solved if at least one sample passes all test cases. Beyond \textit{Pass@k}, we introduce two complementary metrics specifically designed for the characteristics of DL kernels: \textit{Compilation@k} and \textit{$SpeedUp_\alpha@k$}. \textit{Compilation@k} captures the proportion of problems for which at least one of the $k$ generated samples compiles successfully. \textit{$SpeedUp_\alpha@k$} measures runtime performance by reporting the fraction of problems where at least one sample achieves a speedup of at least $\alpha$ (e.g., 2×) compared to a predefined baseline implementation.

To ensure consistent and reproducible performance measurements, we restrict our evaluation of baseline implementations to PyTorch's eager mode, excluding TorchDynamo and other compilation modes due to their platform-dependent behavior, which could introduce measurement inconsistencies. We report all metrics (\textit{Pass@k}, \textit{Compilation@k}, and \textit{$SpeedUp_\alpha@k$}) for $k = 1$ and $k = 5$. For $k = 1$, we use greedy decoding (temperature $= 0.0$) to evaluate the model's top-predicted solution. For $k = 5$, we employ stochastic sampling with temperature $= 0.2$ and top-$p = 0.95$ to balance diversity and quality in the generated outputs, following established practices in prior code generation benchmarks~\cite{codex,javaBench}.  Following prior work~\cite{ouyang2025kernelbench}, we do not apply stochastic sampling to reasoning models, as they entail significantly higher inference time and computational cost.

% This evaluation setup enables a rigorous assessment of both the model’s deterministic performance and its capacity to generate diverse, correct solutions.

% \noindent\textbf{Prompt Design.} 

% To enable the generation of custom kernel-augmented architectures, we design structured prompts tailored to three platforms: CUDA, Pallas, and AscendC. Each prompt begins with a platform-specific instruction outlining the task objective and the expected output format. When available, the prompt includes a concrete example consisting of an original PyTorch module and its corresponding transformed version with embedded custom kernels. By default, the "add" example—illustrated in Figure~\ref{fig:add_kernels}—is provided. This example serves as an in-context demonstration to showcase the correct syntax and integration pattern. The prompt then appends the target module to be transformed, followed by detailed instructions for the code generation task. For CUDA and Pallas, the transformation involves embedding inline custom operators directly within PyTorch modules. In contrast, the AscendC prompt requires a more structured transformation that includes host code, tiling code, and device code, all aligned with a specific kernel name. By grounding the prompt in both instruction and example, our design supports accurate and platform-aware code generation.

\subsection{RQ1: Performance in Default Setting}\label{sec:rq1}
Table~\ref{tab:llm_kernel_stats} presents the overall performance of the selected LLMs across 285 kernel tasks under the default setting, where each task uses the add task as a one-shot example. For each platform, we report three key metrics—\textit{Compilation@k}, \textit{Pass@k}, and \textit{$SpeedUp_1@k$}—for all selected models. Additionally, we include the total number of passed kernels for each model to provide a holistic view of overall performance. The best results for each metric are highlighted in bold to indicate the top-performing models. Overall, MultiKernelBench proves highly challenging: even the best-performing model, Claude-Sonnet-4, successfully completes only a subset of tasks (164 out of 855) under greedy decoding. While sampling search improves performance, a majority of tasks remain unsolved, underscoring the benchmark’s difficulty.

\begin{table*}[htbp]\small
  \centering
\caption{
  \textbf{CUDA Kernel Generation Accuracy by Category and Model using \textit{add} kernels as one-shot exemplars.} 
  Results are shown for three top-performing LLMs. The final column shows the average Pass@1 accuracy across models.
}
  \label{tab:cuda_kernel_stats_by_category}
    \begin{tabular}{l
                    ccc ccc ccc ccc
                    c}
      \toprule
      \multirow{2}{*}{Categories} 
          & \multicolumn{3}{c}{DeepSeek-R1 (\%)} 
        & \multicolumn{3}{c}{Claude-Sonnet-4 (\%)} 
        
        & \multicolumn{3}{c}{Qwen3-235B (think) (\%)} 
        
        & \multirow{2}{*}{Avg Pass@1 (\%)} \\
      \cmidrule(lr){2-4} \cmidrule(lr){5-7} \cmidrule(lr){8-10} 
        & Comp@1 & Pass@1 & $SU_1@1$
        & Comp@1 & Pass@1 & $SU_1@1$
        & Comp@1 & Pass@1 & $SU_1@1$ \\
      \midrule
Activation& 86.7 & 80.0 & 20.0 & 100.0 & 100.0 & 20.0 & 93.3 & 86.7 & 26.7  & 88.9\\
Broadcast& 90.0 & 80.0 & 20.0 & 100.0 & 90.0 & 0.0 & 100.0 & 80.0 & 0.0  & 83.3\\
Convolution& 58.8 & 32.4 & 5.9 & 100.0 & 0.0 & 0.0 & 58.8 & 8.8 & 0.0  & 13.7\\
Full Architecture& 64.0 & 26.0 & 16.0 & 94.0 & 16.0 & 0.0 & 74.0 & 22.0 & 4.0  & 21.3\\
Fusion& 78.0 & 59.0 & 44.0 & 85.0 & 50.0 & 37.0 & 81.0 & 49.0 & 32.0  & 52.7\\
Loss& 100.0 & 42.9 & 28.6 & 85.7 & 42.9 & 14.3 & 100.0 & 57.1 & 42.9  & 47.6\\
Math& 50.0 & 50.0 & 16.7 & 100.0 & 66.7 & 0.0 & 66.7 & 33.3 & 0.0  & 50.0\\
Matrix Multiply& 88.2 & 76.5 & 5.9 & 100.0 & 70.6 & 5.9 & 94.1 & 76.5 & 5.9  & 74.5\\
Normalization& 75.0 & 75.0 & 37.5 & 87.5 & 87.5 & 37.5 & 100.0 & 62.5 & 50.0  & 75.0\\
Optimizer& 60.0 & 40.0 & 40.0 & 100.0 & 100.0 & 80.0 & 100.0 & 60.0 & 60.0  & 66.7\\
Pooling& 83.3 & 33.3 & 0.0 & 100.0 & 66.7 & 50.0 & 66.7 & 16.7 & 0.0  & 38.9\\
Index& 83.3 & 66.7 & 16.7 & 91.7 & 75.0 & 33.3 & 58.3 & 41.7 & 25.0  & 61.1\\
Resize& 90.0 & 50.0 & 30.0 & 100.0 & 40.0 & 20.0 & 80.0 & 50.0 & 20.0  & 46.7\\
Reduce& 100.0 & 100.0 & 20.0 & 80.0 & 80.0 & 0.0 & 100.0 & 80.0 & 20.0  & 86.7\\
    \bottomrule
  \end{tabular}
\end{table*}

\noindent\textbf{Model Performance Comparison.} Claude-Sonnet-4, one of the strongest coding models, passes the largest number of tasks overall. The two reasoning models, DeepSeek-R1 and Qwen3-235B (with thinking mode enabled), achieve the second-highest and third-highest number of passed kernels, respectively, highlighting the effectiveness of reinforcement learning in enhancing model capabilities. In contrast, Qwen2.5-Coder-32B generates the fewest correct kernels, suggesting that a smaller parameter count may negatively impact performance. Notably, while DeepSeek-R1 performs well on CUDA kernels, it lags behind non-reasoning models on AscendC and Pallas kernels, recording the lowest \textit{Pass@k} scores among all seven models for Pallas. This may be attributed to a combination of DeepSeek-R1’s relatively high hallucination rate~\cite{hallucination-leaderboard} and limited exposure to AscendC and Pallas code in its training corpus.

\find{\textbf{Findings:} (1) MultiKernelBench is a challenging benchmark for testing LLMs. (2) Reasoning-augmented models tend to perform better overall, while smaller models like Qwen-32B underperform, showing the impact of both reasoning ability and model scale.}

\noindent\textbf{Platform Comparison.} Model performance varies notably across platforms, primarily reflecting differences in training corpus coverage. On CUDA, GPT-4o achieves the highest \textit{Compilation@1} rate (97.5\%), indicating strong syntactic alignment with CUDA conventions, while DeepSeek-R1 delivers the best \textit{Pass@1} and \textit{$SpeedUp_1@1$}, suggesting superior optimization capabilities for CUDA code. In contrast, performance drops sharply on AscendC and Pallas for most models. DeepSeek-V3 shows the best AscendC result with a modest \textit{Pass@1} of 2.5\%, while all other models remain below 2.1\%. On the Pallas platform, Claude-Sonnet-4 leads with the highest \textit{Pass@1} (8.4\%) and \textit{Pass@5} (10.5\%), whereas DeepSeek-R1, despite its strong CUDA performance, ranks lowest with only 2.8\% \textit{Pass@1}. These platform-specific discrepancies indicate that model effectiveness is highly correlated with the presence of relevant examples in the training data; the weaker performance on AscendC and Pallas likely stems from their limited representation in the pretraining corpora of current LLMs.

\find{\textbf{Findings:} Model performance is highly platform-dependent, reflecting differences in training corpus coverage across CUDA, AscendC, and Pallas.}

\noindent\textbf{Failure Analysis.} On the CUDA platform, compilation errors are relatively uncommon, likely because CUDA is widely documented and has extensive coverage on the internet, despite its C-like syntax and complexity. However, generating functionally correct kernels remains a significant challenge for LLMs. Even Claude-Sonnet-4, the best-performing model, achieves only a 47.0\% \textit{Pass@1} and 55.8\% \textit{Pass@5}—indicating that many tasks still result in failure. We analyze the failure modes of compiled kernels and find that output mismatch is the dominant runtime error, accounting for 53.0\% of failures. Output shape mismatches contribute 16.1\%, while CUDA runtime errors account for 15.9\%. These results highlight the difficulty LLMs face in reasoning about precise output semantics and data structure alignment, even in well-documented domains like CUDA.

\begin{figure}
  \includegraphics[width=\linewidth]{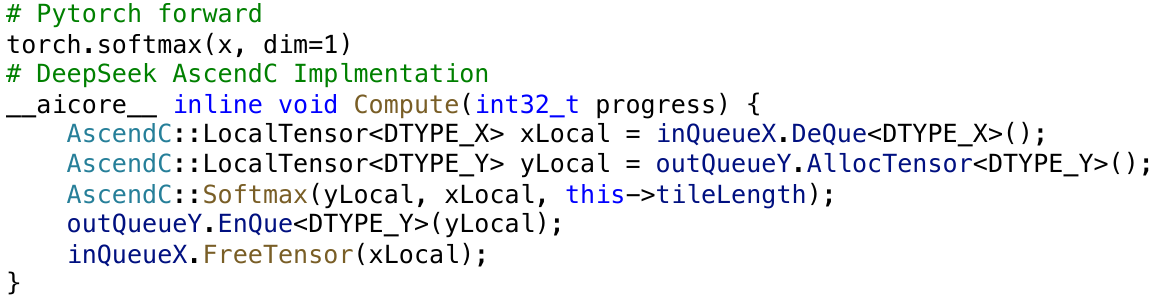}
  \caption{AscendC Compilation Error Example.}
  \label{fig:ascendc_fail}
\end{figure}

Kernels targeting the AscendC platform show the highest rate of compilation errors, mainly due to LLMs’ limited understanding of core AscendC concepts such as SIMD computation, data transfer, and memory management—even with an add kernel as a one-shot example. Notably, 74.8\% of compilation failures contain the keyword "no member named," indicating unfamiliarity with AscendC API usage or attempts to avoid hardware-specific coding. As illustrated in Figure~\ref{fig:ascendc_fail}, for tasks like softmax, LLM generated code directly calls non-existent functions AscendC::Softmax, resulting in compilation errors. This reflects insufficient AscendC code exposure in the pretraining corpus, limiting the models' ability to produce valid code for this platform.

Regarding the Pallas platform, although it also suffers from limited publicly available data, it is built on Python, which shifts many of the compilation-like errors typical in C/C++ to runtime errors instead. Despite a relatively high \textit{Comp@1}, a closer inspection of the top-performing model, Qwen3-235B, reveals a \textit{Pass@5} of only 6.3\%. Failure logs show that the most common runtime exception is \textit{Unexpected Keyword Argument} (22.4\%), indicating that LLMs often hallucinate non-existent keywords for Pallas APIs. Another frequent failure mode is the error \textit{"The Pallas TPU lowering currently supports only blocks of rank >= 1"}, accounting for 5\% of failures. This is likely due to Pallas currently lacking support for scalar values as inputs, outputs, or intermediate computations.
% In several cases, the LLM generated code using arguments that were deprecated in newer library versions but valid in older ones. For example, version 0.4.31 of JAX removed the \texttt{in\_speces\_tree} field from \texttt{jax.experimental.pallas.GridSpec}~\cite{pallas-changelog}, yet the LLM still produced code including this field. This highlights the need for methods to mitigate deprecated API usage in LLM-generated code~\cite{Deprecated-API-usage}.

\find{\textbf{Findings:} (1) LLMs struggle with functional correctness, even on well-documented platforms like CUDA. (2) High compilation failure rates on AscendC suggest a limited understanding of platform-specific syntax. (3) Pallas is prone to runtime errors, driven by incorrect API usage.}

\begin{table*}[htbp]\small
  \centering
  \caption{AscendC Kernel Generation Accuracy by Category and Model using \textit{add} kernels as one-shot exemplars.}
  \label{tab:ascendc_kernel_stats_by_category}
    \begin{tabular}{l
                    ccc ccc ccc ccc
                    c}
      \toprule
      \multirow{2}{*}{Categories} 
        & \multicolumn{3}{c}{DeepSeek-V3 (\%)} 
        & \multicolumn{3}{c}{Claude-Sonnet-4  (\%)} 
        & \multicolumn{3}{c}{Qwen2.5-Coder-32B (\%)} 
        & \multirow{2}{*}{Avg Pass@1 (\%)} \\
      \cmidrule(lr){2-4} \cmidrule(lr){5-7} \cmidrule(lr){8-10} 
        & Comp@1 & Pass@1 & $SU_1@1$
        & Comp@1 & Pass@1 & $SU_1@1$
        & Comp@1 & Pass@1 & $SU_1@1$ \\
      \midrule
Activation& 33.3 & 33.3 & 6.7 & 53.3 & 40.0 & 6.7 & 33.3 & 33.3 & 13.3  & 35.6\\
Broadcast& 40.0 & 20.0 & 20.0 & 10.0 & 0.0 & 0.0 & 30.0 & 0.0 & 0.0  & 6.7\\
\textit{Others} 
           &  \multicolumn{9}{c}{All Pass@1 = 0.0} & 0.0 \\
    \bottomrule
  \end{tabular}
\end{table*}

\begin{table*}[htbp]\small
  \centering
  \caption{Pallas Kernel Generation Accuracy by Category and Model using \textit{add} kernels as one-shot exemplars.}
  \label{tab:pallas_kernel_stats_by_category}
    \begin{tabular}{l
                    ccc ccc ccc ccc
                    c}
      \toprule
      \multirow{2}{*}{Categories} 
        & \multicolumn{3}{c}{Claude-Sonnet-4 (\%)} 
        & \multicolumn{3}{c}{Qwen3-235B (think) (\%)} 
        & \multicolumn{3}{c}{GPT-4o (\%)} 
        & \multirow{2}{*}{Avg Pass@1 (\%)} \\
      \cmidrule(lr){2-4} \cmidrule(lr){5-7} \cmidrule(lr){8-10} 
        & Comp@1 & Pass@1 & $SU_1@1$
        & Comp@1 & Pass@1 & $SU_1@1$
        & Comp@1 & Pass@1 & $SU_1@1$ \\
      \midrule
Activation& 100.0 & 66.7 & 60.0 & 100.0 & 66.7 & 66.7 & 93.3 & 53.3 & 26.7  & 62.2\\
Broadcast& 100.0 & 40.0 & 40.0 & 90.0 & 30.0 & 30.0 & 100.0 & 40.0 & 20.0  & 36.7\\
Full Architecture& 98.0 & 4.0 & 4.0 & 92.0 & 4.0 & 4.0 & 98.0 & 6.0 & 2.0  & 4.7\\
Fusion& 100.0 & 6.0 & 5.0 & 91.0 & 5.0 & 4.0 & 96.0 & 1.0 & 0.0  & 4.0\\
Loss& 100.0 & 0.0 & 0.0 & 85.7 & 0.0 & 0.0 & 100.0 & 14.3 & 14.3  & 4.8\\
Normalization& 100.0 & 25.0 & 25.0 & 100.0 & 12.5 & 12.5 & 100.0 & 12.5 & 0.0  & 16.7\\
    \textit{Others} 
           &  \multicolumn{9}{c}{All Pass@1 = 0.0} & 0.0 \\
    
    \bottomrule
  \end{tabular}
\end{table*}

\subsection{RQ2: Category-wise Analysis}\label{sec:rq2}
Table~\ref{tab:cuda_kernel_stats_by_category}, Table~\ref{tab:ascendc_kernel_stats_by_category}, and Table~\ref{tab:pallas_kernel_stats_by_category} present category-wise results for the CUDA, AscendC, and Pallas backends, respectively, using three metrics under greedy decoding: \textit{Compilation@1}, \textit{Pass@1}, and \textit{$SpeedUp_1@1$}. Due to space limitations, we report results for the top three performing models only. To reflect task difficulty, we also report the average \textit{Pass@1} across these models for each category. Categories in which all models achieved zero \textit{Pass@1} were merged, as they indicate tasks where no correct solution was generated.

\noindent\textbf{Category Comparison. }For CUDA kernels, the top three models achieve the highest \textit{Pass@1} rates in the Activation, Reduce, and Broadcast categories, reaching 88.9\%, 86.7\%, and 83.3\%, respectively. In contrast, the Convolution and Full Architecture categories exhibit significantly lower pass rates—13.7\% and 21.3\%, respectively—highlighting the increased complexity of these tasks. Notably, prior work~\cite{ouyang2025kernelbench} classifies Convolution and Activation as having comparable difficulty, yet their actual \textit{Pass@1} rates diverge markedly (13.7\% vs. 88.9\%), reinforcing the need for our proposed fine-grained categorization. Pooling tasks also show relatively low performance, with a \textit{Pass@1} rate of 38.9\%. Overall, this category-wise analysis reveals substantial variation in model performance across different task types.

On the AscendC and Pallas platforms, LLM-generated kernels succeed in only a few categories. Activation remains the easiest category across all three platforms, likely due to its relatively simple mathematical structure. In contrast, tasks like Reduce and Matrix Multiply—while performing well on CUDA—have a \textit{Pass@1} of zero on AscendC and Pallas, suggesting platform-specific challenges and differences in kernel generation preferences.

\find{\textbf{Findings:} (1) LLMs show strong performance on simple categories like Activation, but struggle with complex tasks such as Convolution, revealing large variation in task difficulty. (2) High-performing categories on CUDA (e.g., Reduce, Matrix Multiply) fail completely on AscendC and Pallas, indicating platform-specific limitations in generalizing kernel generation.}

\begin{figure}[t]
  \includegraphics[width=\linewidth]{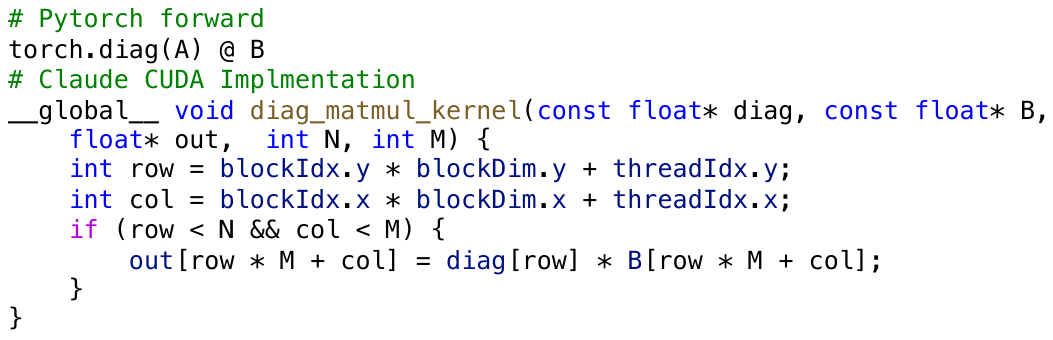}
  \caption{Diagonal Matrix Multiplication.}
  \label{fig:diagonal}
\end{figure}

\begin{figure}[t]
  \includegraphics[width=\linewidth]{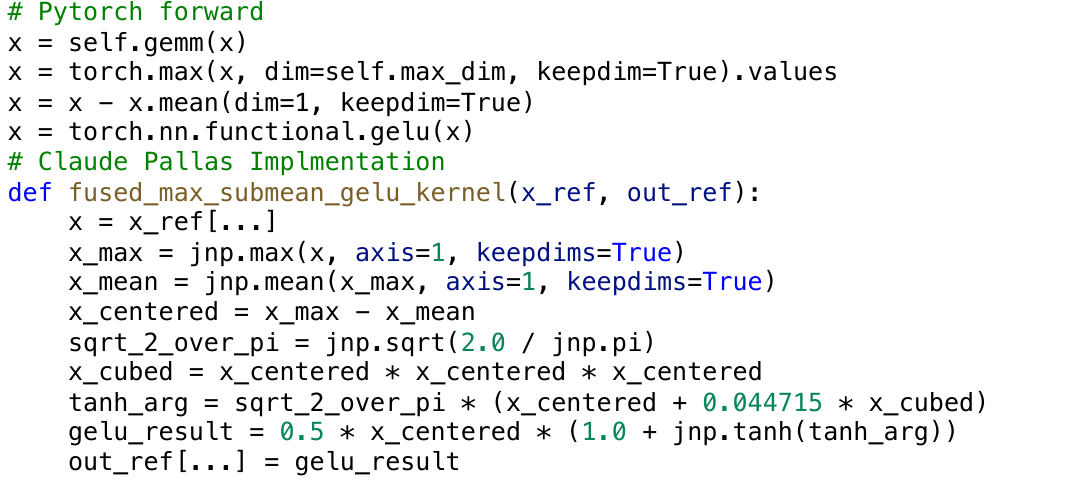}
  \caption{Kernel Fusion Example.}
  \label{fig:fusion}
\end{figure}

\noindent\textbf{Speed Up Cases Analysis.} Analyzing scenarios where LLM-generated kernels outperform native PyTorch kernels reveals several interesting patterns. We identify three common situations where speed-ups occur:
(1) Simple operators, particularly in the activations category: These operations are extremely lightweight (e.g., around 2µs execution time). In such cases, LLM-generated kernels may be slightly faster than PyTorch's native implementations, possibly due to measurement noise or reduced internal overhead. 
(2) Task-specific optimization: LLMs can occasionally discover more efficient strategies tailored to specific tasks. For instance, as shown in Figure~\ref{fig:diagonal}, in diagonal matrix multiplication, the LLM leverages sparsity to skip redundant computations found in general-purpose routines.
(3) Kernel fusion, particularly in fusion tasks: LLM-generated kernels may implicitly combine multiple operations, reducing intermediate memory allocations and data transfers. As illustrated in Figure~\ref{fig:fusion}, for a kernel involving five operations—gemm, max, minus, mean, gelu—the LLM-generated Pallas kernel fuses the latter four, improving performance by minimizing memory bandwidth usage and kernel launch overhead.

\find{\textbf{Findings:} (1) LLMs occasionally discover task-specific optimizations, such as exploiting sparsity in diagonal matrix multiplication. (2) Kernel fusion in LLM-generated code can lead to significant speed-ups by minimizing memory transfers and launch overhead.}

\begin{table*}[htbp]\small
  \centering
  \caption{AscendC kernel generation accuracy by category and model using \emph{in-category} one-shot examples.
  $\Delta$Pass indicates the absolute improvement in Pass@1 over the baseline setting using an \textit{add} kernel as the one-shot example. 
  The last row summarizes the relative improvement in overall.}
  \label{tab:ascendc_in_category}
  \begin{tabular}{l
                  ccc c
                  ccc c
                  ccc c}
    \toprule
    \multirow{2}{*}{Category}
      & \multicolumn{4}{c}{Claude-Sonnet-4 (\%)}
      & \multicolumn{4}{c}{GPT-4o (\%)}
      & \multicolumn{4}{c}{DeepSeek-V3 (\%)} \\
    \cmidrule(lr){2-5} \cmidrule(lr){6-9} \cmidrule(lr){10-13}
        & Comp@1 & Pass@1 & $SU_1@1$ & $\Delta Pass$
        & Comp@1 & Pass@1 & $SU_1@1$ & $\Delta Pass$
        & Comp@1 & Pass@1 & $SU_1@1$ & $\Delta Pass$ \\
    \midrule
Activation      &60.0 & 60.0 & 20.0  & +20.0    
                & 46.7 & 40.0 & 13.3 & +6.7 
               & 40.0 & 40.0 & 6.7  & +6.7 
                \\
Loss           & 14.3 & 14.3 & 14.3 & +14.3
               & 28.6 & 14.3 & 14.3 & +14.3 
               & 0.0  & 0.0  & 0.0  & +0.0 
                 \\
Matrix Multiply & 52.9 & 23.5 & 0.0 & +23.5   
               & 82.4 & 35.3 & 0.0  & +35.3 
               & 52.9 & 23.5 & 0.  & +23.5  \\
Normalization  & 62.5 & 0.0 & 0.0 & +0.0
               & 12.5 & 0.0 & 0.0  & +0.0 
               & 25.0 & 0.0 & 0.0  & +0.0 
                \\
Reduce         & 0.0  & 0.0  & 0.0  & +0.0 
               & 0.0  & 0.0  & 0.0  & +0.0 
               & 0.0  & 0.0  & 0.0  & +0.0 \\
\midrule
\textbf{Rel. Improve} 
                & \multicolumn{1}{c}{\textbf{↑200.0}} & \textbf{↑133.3} & \textbf{↑300.0} &
               & \multicolumn{1}{c}{\textbf{↑380.0}} & \textbf{↑160.0} & \textbf{↑200.0} & 
               & \multicolumn{1}{c}{\textbf{↑240.0}} & \textbf{↑100.0} & \textbf{↑0.0}
               \\
    \bottomrule
  \end{tabular}
\end{table*}

\begin{table*}[htbp]\small
  \centering
  \caption{Pallas kernel generation accuracy by category and model using \emph{in-category} one-shot examples.}
  \label{tab:pallas_in_category}
  \begin{tabular}{l
                  ccc c
                  ccc c
                  ccc c}
    \toprule
    \multirow{2}{*}{Category} 
      & \multicolumn{4}{c}{Qwen3-235B (think) (\%)}
      & \multicolumn{4}{c}{DeepSeek-V3 (\%)} 
      & \multicolumn{4}{c}{Qwen3-235B (\%)} \\
    \cmidrule(lr){2-5} \cmidrule(lr){6-9} \cmidrule(lr){10-13}
        & Comp@1 & Pass@1 & $SU_1@1$ & $\Delta Pass$
        & Comp@1 & Pass@1 & $SU_1@1$ & $\Delta Pass$
        & Comp@1 & Pass@1 & $SU_1@1$ & $\Delta Pass$ \\
    \midrule
Activation     & 100.0 & 73.3 & 40.0 & +6.6
               & 100.0 & 86.7 & 46.7 & +33.4 
               & 93.3 & 66.7 & 40.0  & +6.6 \\
Loss           & 100.0 & 42.9 & 14.3 & +42.9
               & 100.0 & 14.3 & 0.0  & +14.3 
               & 100.0 & 28.6 & 0.0 & +28.5 \\
Matrix Multiply         & 100.0 & 41.2 & 0.0  & +41.2
               & 94.1 & 35.3 & 0.0 & +35.3 
               & 100.0 & 35.3 & 0.0  & +35.3 \\
Normalization  & 100.0 & 25.0 & 12.5  & +12.5 
               & 75.0 & 12.5 & 0.0  & +0.0 
               & 100.0 & 25.0 & 12.5  & +25.0 \\
Reduce         & 100.0 & 80.0 & 40.0  & +80.0 
               & 100.0 & 40.0 & 20.0  & +40.0 
               & 100.0 & 20.0 & 0.0  & +20.0 \\
\midrule
\textbf{Rel. Improve} 
               & \multicolumn{1}{c}{\textbf{↑2.0}} & \textbf{↑145.5} & \textbf{↑0.0}&
               & \multicolumn{1}{c}{\textbf{↑16.7}} & \textbf{↑155.6} & \textbf{↑166.7}&
               & \multicolumn{1}{c}{\textbf{↑0.0}} & \textbf{↑133.3} & \textbf{↑16.7} & \\
    \bottomrule
  \end{tabular}
\end{table*}

\subsection{RQ3: Category-Aware One Shot}
Experiments in RQ1 and RQ2 use add kernels as one-shot examples and demonstrate very low pass rates on the two platforms: AscendC and Pallas. This is likely due to the limited presence of these two languages in the training data of LLMs. We argue that while an add kernel can offer basic syntax and usage patterns, it fails to convey the domain-specific knowledge needed for more complex or specialized tasks. Each category typically requires distinct domain knowledge and coding conventions. For example, on Huawei NPUs, matrix multiplication and general math operations use different computation units and follow different programming paradigms. Therefore, in this research question, we explore whether providing one-shot examples from the same category as the target task can help LLMs better capture category-specific structures and improve kernel generation performance.

For AscendC and Pallas, we refer to official documentation~\cite{pallas-kernel,ascendc-document} and open-source repositories~\cite{ascendc-samples} to collect example implementations. Specifically, we gather one representative example for five categories on each platform: Activation, Loss, Matrix Multiply, Normalization, and Reduce. These categories are chosen because suitable one-shot examples are available, with clear and concise implementations. We use the same prompt template as in RQ1 and RQ2, replacing the add task with a category-specific one-shot example.

Table~\ref{tab:ascendc_in_category} and Table~\ref{tab:pallas_in_category} present the results of using this category-aware one-shot strategy. Due to space limitations, we report results only for the top three models with the highest pass rates under this setting. In addition to the standard three metrics per category—Comp@1, Pass@1, and $SU_1@1$—we also report improvements over the default (add) setting. For each model, we include an additional column showing the absolute increase in Pass@1 (i.e., new value minus baseline). Additionally, the last row summarizes the average relative improvement for each metric, calculated as the percentage increase over the baseline values.

For AscendC kernels, using category-aware one-shot examples significantly improves both compilation and execution success rates. For example, GPT-4o achieves a 380\% relative improvement in compilation success rate and a 160\% improvement in correctness (Pass@1) compared to the baseline. At the category level, the most notable gains occur in Matrix Multiply: under the default setting, none of the models could produce correct kernels, but with category-aware one-shot prompting, they begin to generate valid outputs—though still without achieving runtime speedup. These results highlight that in-category examples provide valuable domain knowledge for LLMs, helping them better understand hardware-specific programming patterns and constraints, ultimately improving both compilation and correctness outcomes.

% This suggests that LLMs can effectively extract essential domain knowledge from in-category examples, leading to improvements in both compilation and correctness.

Despite overall progress, LLMs still struggle to generate correct AscendC kernels for certain categories such as Normalization and Reduce. In the case of Reduce, our analysis shows that many generated kernels incorrectly handle 3D input tensors, while the one-shot example only uses a 2D tensor. This highlights a broader limitation in shape-specific reasoning. These observations suggest the need for more fine-grained categorization—potentially based on input tensor shapes—and the development of shape-aware prompting strategies to better guide LLMs in handling such cases accurately.

For Pallas kernels, pass rates improve substantially, with relative gains exceeding 100\% compared to the baseline. This indicates that well-chosen in-category examples play a crucial role in guiding LLMs toward correct kernel generation. Unlike AscendC, every category shows improvement on Pallas. This difference may be due to the simplicity of the Pallas language: most example programs are much shorter than their AscendC counterparts and delegate low-level, hardware-specific optimizations to the underlying compiler. These findings suggest that designing high-level domain-specific languages that abstract away low-level implementation details can help LLMs generate correct and efficient code more easily, enhancing the usability of such languages for AI-assisted programming.

 % suggesting that LLMs can more effectively generalize when provided with Pallas kernel examples.

\find{\textbf{Findings: }(1) Category-aware one-shot prompting significantly boosts LLM performance on platforms with limited training exposure. (2) On Pallas, all categories see substantial performance gains, likely due to its concise syntax and compiler-managed optimizations.}

\section{Threat to Validity}
The first threat to validity lies in the construction of our benchmark, including both the reference implementations and example kernels. To mitigate this, we manually reviewed all added kernels and verified their correctness through compilation and execution tests.
A second threat concerns the generalizability of our findings to other LLMs. To address this, we selected a diverse set of models, ranging from smaller-scale (32B) to very large-scale models (681B), and including both open-source (DeepSeek, Qwen) and proprietary models (GPT-4o).
Another threat is whether our results generalize to other hardware platforms and programming languages. As the landscape of DL accelerators continues to expand—with platforms like AMD (HIP) and more general frameworks like Triton—this is a relevant concern. To facilitate extensibility, our benchmark includes a backend abstraction layer that allows new platforms or languages to be integrated with minimal effort. For instance, supporting Triton requires fewer than 20 lines of additional code. We plan to incorporate such platforms in future work.
\section{Related work}
\subsection{LLM-based Code Generation}

Large Language Models (LLMs) have demonstrated impressive capabilities in generating source code from natural language prompts. General-purpose models such as the DeepSeek series~\cite{deepseekai2024deepseekv2strongeconomicalefficient,deepseekai2025deepseekv3technicalreport}, LLaMA series~\cite{grattafiori2024llama3herdmodels, touvron2023llamaopenefficientfoundation, touvron2023llama2openfoundation}, and GPT series~\cite{gpt3,openai2024gpt4technicalreport} have all shown strong performance on code generation tasks. These models learn programming patterns, syntax, and problem-solving strategies by training on large-scale datasets that combine code and natural language.

To further enhance code generation capabilities, many approaches apply continual pretraining and code-specific instruction tuning to strong foundation models. For example, Code Llama~\cite{code_llama} and DeepSeek-Coder-V2~\cite{deepseekai2024deepseekcoderv2breakingbarrierclosedsource} are built by further training LLaMA 2 and DeepSeek-V2 respectively, using trillions of additional tokens from public GitHub repositories and related text. Following continual pretraining, instruction tuning is often used to improve task alignment. Models such as WizardCoder~\cite{luo2024wizardcoder}, Magicoder~\cite{Magicoder}, and DataScope~\cite{DataScope} utilize synthetic instruction data tailored for coding tasks, leading to further improvements.

Despite these advances, challenges remain—especially in generating correct code for tasks requiring deep context or domain-specific APIs. Retrieval-augmented generation (RAG) has emerged as a promising direction to address these issues. For instance, RePoCoder~\cite{zhang2023repocoder} integrates a similarity-based retriever with a pretrained code LLM in an iterative retrieval-generation pipeline to support repository-level code completion. Similarly, the Repository-Level Prompt Generator~\cite{pmlr-Repository-Level} generates example-specific prompts by selecting useful code snippets to inject domain knowledge into the prompt. DocPrompting~\cite{zhou2023docprompting} enhances code generation by retrieving relevant documentation snippets.

\subsection{Benchmarks for Code Generation}

Evaluating code generation models requires diverse and rigorous benchmarks that assess both functional correctness and code quality across varying levels of abstraction. At the function level, datasets like HumanEval~\cite{codex} and MBPP~\cite{mbpp} pioneered this space by verifying the correctness of generated code using unit tests. These benchmarks remain widely used for evaluating LLM performance on short, self-contained programming problems. To explore more complex generation scenarios, later benchmarks expanded to higher levels of granularity: ClassEval~\cite{ClassEval} evaluates generation at the class level, while DevEval~\cite{li-etal-2024-deveval} and RepoEval~\cite{zhang2023repocoder} assess model performance on repository-level tasks, including long-context reasoning and multi-file consistency.

While most early benchmarks focus on general-purpose coding tasks—such as algorithm design—more recent efforts have shifted toward domain-specific or library-intensive code generation. For example, ODEX~\cite{ODEX} and TorchDataEval~\cite{Private-Library} evaluate a model’s ability to use specialized libraries for scientific computing or data loading. DOMAINEVAL~\cite{DOMAINEVAL} introduces an automated benchmark construction pipeline that targets domain-specific tasks such as cryptography and systems programming. Similarly, BigCodeBench~\cite{zhuo2025bigcodebench} evaluates models on high-level, multi-step tasks like data analysis and web development, which require the integration of diverse APIs and function calls, resembling real-world software engineering.

A particularly challenging and high-impact domain involves generating hardware-efficient kernels, where models must not only produce functionally correct code but also achieve high runtime performance. Benchmarks like KernelBench~\cite{ouyang2025kernelbench} and TritonBench~\cite{li2025tritonbenchbenchmarkinglargelanguage} focus on this task: KernelBench includes 250 kernel tasks and uses the fast@p metric to jointly assess correctness and speedup; TritonBench assesses 184 real-world Triton operators with a focus on functional and performance profiling on NVIDIA GPUs. However, both are constrained to the NVIDIA ecosystem, overlooking other important platforms. To address this, MultiKernelBench introduces a more comprehensive, platform-agnostic evaluation framework by supporting multiple backends.

\section{Conclusion}
As demand grows for efficient DL kernel development across a diverse range of hardware platforms, leveraging LLM for automatic kernel generation has become increasingly vital. Yet, current benchmarks lack the breadth, granularity, and extensibility needed for rigorous evaluation. In this work, we present \emph{MultiKernelBench}, the first comprehensive, multi-platform benchmark suite for DL kernel generation. It spans GPUs, NPUs, and TPUs, and includes 285 tasks across 14 functional categories. Through experiments with seven diverse LLMs, we observe substantial performance variability across both platforms and kernel categories. Notably, we show that a simple, category-aware prompting strategy significantly improves generation performance on AscendC and Pallas. Overall, MultiKernelBench offers a realistic and extensible foundation for benchmarking and advancing LLMs in DL system programming, facilitating future research in kernel generation and optimization.
\bibliographystyle{ACM-Reference-Format}
\bibliography{reference}

\end{document}